\documentclass[11pt,twoside]{article}

\usepackage{asp2006}
\usepackage{epsf}
\usepackage{psfig}
\usepackage{lscape}

\begin{document}
\title{Radio properties of Very High Energy  gamma-ray sources
}   
\author{J.M. Paredes}   
\affil{Departament d'Astronomia i Meteorologia and Institut de Ci\`encies del 
Cosmos (ICC), Universitat de Barcelona (UB/IEEC), Mart\'{\i} i Franqu\`es 1, 
08028 Barcelona, Spain (email: jmparedes@ub.edu) \\
}    

\begin{abstract} 
Radio observations of very high energy (VHE) gamma-ray sources are fundamental to identify and 
reveal their nature, as well as to understand the physics behind these energetic 
sources. I will comment on some characteristics of extragalactic sources detected 
at TeV energies and radio wavelengths, and galactic sources such as gamma-ray 
binaries, supernova remnants (SNRs), or pulsar wind nebulae (PWNs). Special emphasis will be put on unidentified extended 
TeV sources,  in which deep radio observations can severely constrain the 
proposed models and shed light on the possible sources powering the TeV emission.

\end{abstract}


\section{Introduction}   

The identified very high energy (VHE) gamma-ray sources belong to the classes
of active galactic nuclei (AGN), pulsar wind nebula (PWN), shell-type supernova remnants (SNR), X-ray
binaries (XRB) and, as reported recently, young stellar clusters (Aharonian et al.
2007a). In addition to their TeV emission, all of them are also X-ray and/or
radio emitters, as it is expected from the theoretical point of view. The emission of VHE
gamma-rays is explained by the presence of relativistic electrons or protons.
In leptonic scenarios, gamma-rays are produced by IC scattering of background
photons by high-energy electrons. These electrons should produce also
significant X-ray and radio emission through the synchrotron mechanism. In the
hadronic scenarios, proton interactions would produce charged pions and
ultimately secondary electrons that should produce also X-ray and radio
emission. There are, however, $\sim$ 10--30 TeV extended sources ($\sim$
0.05--0.3$^{\circ}$) that lack any plausible radio/X-ray counterparts and
remain unidentified. Some TeV unidentified sources have strong upper limits in
radio with a TeV to radio flux ratio of $F_{\rm TeV}/F_{\rm radio}>10^3$
(Atoyan et al. 2006). These TeV sources with unknown counterpart at lower
energies seem to indicate that either we are dealing with a new class of
objects or that sufficiently deep X-ray and radio observations have not yet
been made. The study at X-ray and radio wavelengths of these unidentified
sources can put significant constraints on the source physics and particle
acceleration processes. 

The instrumentation developed in the last decade in the field of VHE allows us  to conduct surveys, morphological and
spectral studies, detect faint and diffuse emission, detect variability, or conduct phase-resolved spectroscopy
of variable sources. However, to identify and understand the physics of the VHE sources it is necessary to observe them at
other wavelengths, mainly radio and X-rays. Although most of the detected VHE sources known have been identified with well
known sources at other wavelengths (AGNs, SNRs, X-ray binaries, etc), a significant part of them remain unidentified. \\

\section{Extragalactic sources}    

Most of the extragalactic objects detected at TeV energies are AGNs of the type of high frequency peaked BL Lacertae objects
(HBL). Synchrotron radiation from radio to X-rays is produced by relativistic electrons in the jet in presence of magnetic
fields. The inverse Compton (IC) scattering of these electrons with ambient photons can produce VHE gamma-ray emission. The
ambient photons can be of synchrotron origin (Synchrotron Self Compton) or external (External Compton). The two emitting
processes, synchrotron and IC, produce an spectral energy distribution with two characteristic bumps. On the other hand,
hadronic processes could
also be at work, and the question about the nature of the emitting particles remains open (see, e.g., B\"ottcher 2007 and 
references therein). So far, more than 20 of these objects have been detected, and rapid flares have been 
discovered in Mrk~501 (Albert
et al. 2007a) and PKS~2155$-$034 (Aharonian et al. 2007b). In Mrk~501 the flux was doubled in timescales down to 2 minutes and
there was a delay between the peaks of F($<$0.25~TeV) and F($>$1.2~TeV), which could be interpreted as a progressive
acceleration of electrons in the emitting plasma blob or to be the result of a vacuum refractive index in the context of
quantum gravity (Albert et al. 2007b). In  PKS~2155$-$034, the flux doubled in timescales down to 3 minutes, implying 
Doppler
factors above 100 (Aharonian et al. 2007c). 

We note that the detection by H.E.S.S. of the radio galaxy M87 (Aharonian et al. 2006a), previously detected by HEGRA,
confirm that TeV gamma-ray emission can be produced by extragalactic sources other than blazars. M87 also shows fast
variability,  compatible with an emitting region with the size of the Schwarschild radius of the central black hole 
(Albert et al. 2008a).

\section{Galactic sources}  

The VHE galactic sky is composed mainly by three astronomical populations: supernova remnants (SNR), pulsar wind nebulae
(PWN) and X-ray binary stars. 

About a dozen of SNR have been detected with the Cherenkov telescopes, and few of them display a TeV shell that follows
the synchrotron keV emission seen at X-ray (e.g., RX~J1713.7$-$3946, Aharonian et al. 2004). This behaviour agrees with a
leptonic scenario, where the X-ray emission is synchrotron radiation produced by relativistic electrons and the TeV emission
is produced by IC up-scatter ambient photons. The TeV spectra, however, is better explained within the hadronic scenario  in
which VHE gamma-rays are produced by proton-proton interactions and neutral pion decay.  

PWN is the class of galactic astronomical objects with the largest number of members (about 20). They are TeV extended
sources showing a morphology often asymmetric, where the centre of gravity of the TeV emission does not coincide with the
position of the pulsar. In some cases, the offset can be produced by the high peculiar 
velocity of the pulsar whereas in other cases, such as
Vela X, it could be produced by the propagation of the reverse shock formed at the termination of the pulsar wind in an
inhomogeneous medium (Gaensler \& Slane 2006)

In general, the TeV emission in PWN can be explained in a leptonic scenario. The electrons of the relativistic wind of the
pulsar produce radio and X-ray radiation, whereas IC scattering of photons from the CMB or from nearby stars produces
TeV emission. Under this scenario, the larger size observed at TeV energies of HESS~J1825$-$137 when compared with the
observed X-ray size, can be explained by the shorter radiating loss time of the electrons producing synchrotron X-rays than
those radiating VHE gamma-rays by IC scattering (Aharonian et al. 2008a).

An interesting example of composite PWN/SNR is G~0.9+0.1, detected at VHE energies (Aharonian et al. 2005a). The
emission can be explained by IC scattering of relativistic electrons and originated in the plerionic core.

At present,  there are four X-ray binaries that have been detected at TeV energies. All of them
have a bright high mass primary star, which provides a huge UV photon field for inverse Compton scattering. 
Three of them, PSR~B1259$-$63 (Aharonian
et al. 2005b), LS~I~+61~303 (Albert et al. 2006) and LS~5039 (Aharonian et al. 2005c) have been detected in several  parts of
their orbits and show a variable TeV emission. The other source, Cygnus~X-1, has been detected once during a  flare (Albert
et al. 2007c). LS~I~+61~303 shares with LS~5039 the quality of being the only two known high-energy emitting X-ray binaries
that are spatially coincident with sources above 100~MeV listed in the Third EGRET catalog (Hartman et al. 1999). 
The compact
secondary star is a black hole in the case of Cygnus~X-1 and a neutron star in the case of PSR~B1259$-$63. For LS~I~+61~303
and LS~5039 there are not yet evidences supporting the black hole or neutron star nature of the compact object.

The four sources are also radio emitters. 
Cygnus~X-1 displays a $\sim$15~mJy and
flat spectrum relativistic compact (and one-sided) jet ($v>0.6c$) during the
low/hard state (Stirling et al. 2001). Also, arc-minute extended radio emission around 
Cygnus~X-1 was found (Mart\'{\i} et al. 1996) using the VLA. Their disposition reminded 
of an elliptical ring-like shell with Cygnus~X-1 offset from the center. 
Later, as reported in Gallo et al. (2005), 
such structure was recognised as a
jet-blown ring around Cygnus~X-1. 
This ring could be the result of a strong shock
that develops at the location where the pressure exerted by the collimated jet,
detected at milliarcsec scales, is balanced by the ISM. The observed thermal 
Bremsstrahlung radiation would be produced by the ionized gas behind the bow shock. 

PSR~B1259$-$63 contains a 47.7 ms radio pulsar orbiting  around its massive companion  every 3.4 years in a very
eccentric orbit. The radiation mechanisms and  interaction geometry in this pulsar/Be star system was studied in Tavani \&
Arons (1997). In a hadronic scenario, the TeV  light-curve, and radio/X-ray light-curves, can be produced by the collisions
of high energy protons accelerated by the pulsar wind and the circumstellar disk (Neronov \& Chernyakova 2007). A very
different model is presented in Khangulyan et al. (2007), where it is shown that the TeV light curve can be explained by  IC
scenarios of gamma-ray production.

LS~I~+61~303 shows periodic non-thermal radio outbursts on average every $P_{\rm orb}$=26.4960~d (Taylor \& Gregory 1982).
Massi et al. (2004) reported the discovery of an extended jet-like and apparently precessing radio emitting structure at
angular extensions of 10--50~milliarcseconds. VLBA images obtained during a full orbital cycle show a rotating
elongated morphology (Dhawan et al. 2006), which may be consistent with a model based on the interaction between the
relativistic wind of a young non-accreting pulsar and the wind of the stellar companion (Dubus 2006; see nevertheless 
Romero et al. 2007 for a rebuttal of this scenario).

The radio emission of LS~5039 is persistent, non-thermal and variable but no strong radio outbursts or periodic variability
have been detected so far (Rib\'o et al. 1999, 2002). VLBA observations allowed the detection of an elongated radio
structure, interpreted as relativistic jets  (Paredes et al. 2000). The discovery of this bipolar radio structure, and the
fact that LS~5039 was the only source in the field of the EGRET source 3EG~J1824$-$1514 showing X-ray and radio emission,
allowed to propose the physical association of both sources (Paredes et al. 2000). High resolution (VLBI) radio observations
of LS~5039 along the orbit are necessary for the detection of morphological and astrometric changes, which can be useful to
disantangle the nature of the compact source {(Rib\'o et al. 2008). A theoretical discussion of the radio properties of
LS~5039 can be found in Bosch-Ramon (2009).}

Some properties of these systems and an individual description of them can be find in Paredes (2008). Also, a thorough 
discussion of the different theoretical models for LS~5039 and LS~I~+61~303 is presented in Bosch-Ramon \& Khangulyan
(2009).

\begin{figure}
\centering \epsfxsize=.50\textwidth \epsfbox{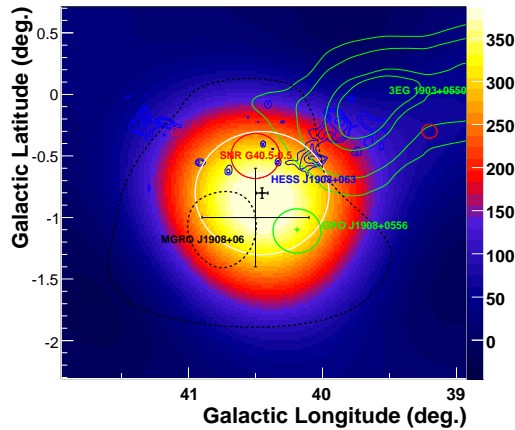} 
\caption{View of the field around the position of HESS~J1908+063 showing the H.E.S.S. excess map. The dotted black line shows the
MILAGRO significance contours (5$\sigma$ and 8$\sigma$, inner and outer). The big black cross represents the center of gravity of the MILAGRO
source. The position of the EGRET GeV source GRO~J1908+0556 is marked with a
green cross as well as the 1$\sigma$ error in the position. The 3EG~1903+0550
contours (99, 95, 68 and 50\% confidence levels) are shown in
green. The red circle marks the size and position of the radio-bright 
SNR~G40.5$-$0.5. Contours in blue show the $^{13}$CO molecular cloud in the 
velocity range between 45 and 65 km~s$^{-1}$. From Djannati-Atai et al.\
(2007).
 }
 \label{hess}
 \end{figure}

\section{Unidentified sources}

Nearly one third of the VHE $\gamma$-ray sources discovered during these last years by the new generation of \^Cerenkov
telescopes remain yet as unidentified. A significant number of them show extended morphologies, making the identification of counterparts at lower energies a very difficult task. The most representative of
this new population of galactic sources is TeV~J2032+4130, discovered with HEGRA in the direction of the Cygnus OB2 star
association (Aharonian et al. 2002). Deep radio observations using the GMRT at 610 MHz and the VLA at 1.4 and 5 GHz covering
the TeV J2032+4130 field have revealed both compact and extended radio sources at arcsecond scales, showing that this
particle accelerator is not so dark at low energies than previously thought (Paredes et al. 2007). It is also worth to note
that a diffuse X-ray counterpart has also recently been  reported thanks to deep {\it XMM-Newton} observations (Horns
et al. 2007). Other examples are HESS~J1303$-$631 (Aharonian et al. 2005d), the second unidentified TeV source, as well as
other unidentified sources found in the HESS galactic plane survey (Aharonian et al. 2005e). This survey of the inner Galaxy
has revealed new sources that remain unidentified and have not been associated with any object from which VHE emission is
expected. A recent study of the environment of eight of these unidentified extended TeV sources with high detection
significance revealed no plausible counterpart (Aharonian et al. 2008b), although one of them, HESS~J1731$-$347, is
associated with G353.6$-$0.7, likely an old SNR.

The Milagro Gamma Ray Observatory, an extensive air shower array sensitive to multi-TeV photons, has detected four sources
(Abdo et al.\ 2007). One of them is the Crab Nebula, which appeared point-like. The other three sources are clearly extended
and are located in the Cygnus/Aquila region of the Galaxy. One of these extended sources, MGRO~J2031+41, is coincident with
the unidentified HEGRA source TeV~J2032+4130, which has recently been confirmed by the detection reported by the MAGIC
Collaboration (Albert et al.\ 2008b). The first new extended source discovered by MILAGRO is MGRO~J2019+37. For this
unidentified TeV source, deep GMRT observations at  610~MHz have detected more than one hundred of new radio sources 
(Paredes et al. 2009). These GMRT observations showed other interesting radio sources which were  not obvious at first
glance, but they become evident when looking in detail at each source in the field; some of them display a resolved
morphology. The second extended new source discovered is MGRO~J1908+06, located in Aquila. Dedicated
observations by H.E.S.S. have recently revealed the detection of a new source from 0.3 to 50~TeV, dubbed
HESS~J1908+063, in a position compatible with the MILAGRO one (Djannati-Atai et al.\ 2007). In the field of view of
HESS~J1908+063 there are several catalogued radio sources (see Fig.~\ref{hess}). One of them is the SNR~G40.5$-$0.5
(Green 2006). If it is associated with the VHE source, the fact that the 22$^{\prime}$ size of the shell is smaller than the
43$^{\prime}$ FWHM of HESS~J1908+063 would contrast to previously discovered HESS sources identified with shell-type VHE
emitters, such as RX~J1713.7$-$3946 or RCW~86 that have the same size at both energies. 

\section{Summary}

The VHE sky is getting more and more crowded as the Cherenkov instruments increase their sensitivity. This has
brought a significant population of sources of unclear physics (e.g. X-ray binaries) or even unknown nature (unidentified TeV
sources). In such a context, radio observations are of primary importance since they probe the behavior of relativistic
particles at energy ranges different from those related to VHE emission, helping to unveil the physics of the sources, and
therefore to solve the misteries opened by very high-energy astronomy.

\acknowledgements 
The author acknowledges support of the Spanish Ministerio de Educaci\'on y 
Ciencia (MEC) under grant AYA2007-68034-C03-01 and FEDER funds. This research has made use
of the NASA's Astrophysics Data System Abstract Service, and of the SIMBAD database, operated at
CDS, Strasbourg, France.


\end{document}